# Ferroelectric Domain Wall Logic Gates

Ahmet Suna[1,#], Conor J. McCluskey[1], Jesi R. Maguire[1], Amit Kumar[1], Raymond G. P. McQuaid[1] and J. Marty Gregg[1,*]

[1] School of Mathematics and Physics, Queen's University Belfast, Belfast, BT7 1NN, U.K.

*email: m.gregg@qub.ac.uk
#email: a.suna@qub.ac.uk

Fundamentally, lithium niobate is an extremely good electrical insulator. However, this can change dramatically when 180° domain walls are present, as they are often found to be strongly conducting. Absolute conductivities depend on the inclination angles of the walls with respect to the [001] polarisation axis and so, if these inclination angles can be altered, then electrical conductivities can be tuned, or even toggled on and off. In 500nm thick z-cut ion-sliced thin films, localised wall angle variations can be controlled by both the sense and magnitude of applied electrical bias. We show that this results in a diode-like charge transport response which is effective for half-wave rectification, albeit only at relatively low ac frequencies. Most importantly, however, we also demonstrate that such domain wall diodes can be used to construct "AND" and inclusive "OR" logic gates, where "0" and "1" output states are clearly distinguishable. Realistic device modelling allows an extrapolation of results for the operation of these domain wall diodes in more complex arrangements and, although non-ideal, output states can still be distinguished even in two-level cascade logic. Although conceptually simple, we believe that our experimental demonstration of operational domain wall-enabled logic gates represents a significant step towards the future broader realisation of "domain wall nanoelectronics".

**Introduction**

While bulk ferroelectrics are generally either insulators or wide band-gap semiconductors, it is now clear that ferroelectric domain walls (interfaces in ferroelectric microstructures that separate differently oriented domains) can show significantly enhanced electrical transport [1-8]. In some materials, such as bismuth ferrite, even though band-gap alterations with respect to bulk might be expected at domain walls [9,10], observed conductivity seems to be primarily extrinsic in origin, controlled by the extent to which walls act as aggregation sites for point defects [11-15]. In other systems, polar discontinuities at so-called "charged walls" lead to intense local fields and the associated enhanced domain wall conduction seems to be more intrinsic in nature. Band-bending is thought to be sufficient to either push the conduction band below the Fermi Level (creating electronic conduction within the wall) or raise the valence band above it (creating hole conduction), depending on the sense of the polarisation discontinuity present (being either "head-to-head" or "tail-to-tail") [5,16,17].

In head-to-head charged 180° domain walls in lithium niobate ($LiNbO_3$ or LNO), a correlation between the size of the polar discontinuity and the observed conductivity has been categorically established [17]. The angles of domain wall inclination with respect to the polarisation axis determine the magnitude of the polar discontinuity and can be actively controlled by applied electric fields, which monotonically change associated conduction levels. In the specific case of z-cut ion-sliced LNO thin film capacitor structures, with thin film AuCr lower electrodes [18-21], 180° head-to-head domain walls, created by partial ferroelectric switching, are initially strongly inclined and highly conducting (see schematic in figure 1a). With positive bias applied through surface electrodes or conducting atomic force microscopy (cAFM) tips, the inclination angles of the walls appear to be maintained or enhanced and strong conductivity remains [18]. However, when a negative bias is applied (or if the films are left in ambient conditions for several days), regions of the walls extending some ~50-100nm beneath the thin film surface "straighten" (better align with the polar axis) and local domain wall conductivity collapses [18, 20] (see schematic in figure 1b). "Straightened" regions break the percolation pathway along the conducting domain wall conduits that bridge the interelectrode gap and thereby create a high resistance state for the device as a whole. This variation in near-surface domain wall inclination under positive and negative bias should be expected to result in a strongly asymmetric current-voltage characteristic, with currents shutting down for negative bias and developing progressively more strongly for positive bias applied to the LNO top surface. In other words, a diode-like response should be expected.

Herein, we present experimental evidence to confirm that domain wall diodes can indeed be generated in z-cut ion-sliced LNO thin films, as a direct result of electric-field control of the subsurface domain wall inclination angle. We also show that alternating current rectification, demonstrated recently in x-cut material [22], can also be seen in these z-cut films. Most importantly, however, we show that LNO domain wall diodes can be successfully used to realise simple "AND" and inclusive "OR" logic gates, which show a clear distinction between "0" and "1" output states. Such domain wall logic gates had recently been envisaged [23] but have not yet been experimentally realised in any prior research. When the experimentally observed characteristics of the domain wall diodes are incorporated into circuit simulation software, we have been able to show that output state distinguishability is maintained even in two-level cascade logic configurations. Viable Boolean logic gates, the operation of which is entirely dependent on the local manipulation of domain wall conductivity, have hence been established, further pushing forward possibilities for domain wall-based nanoelectronics [24].

**Results**

**Electric field control of subsurface domain wall inclination angle.** 500nm thick commercially obtained z-cut ion-sliced single crystal LNO thin film samples, with 150nm thick Cr/Au lower electrodes, were used for this work (made by NanoLN). The polarization direction of the ferroelectric layer in the as-received samples was oriented out-of-plane and perpendicular to the sample surface. However, when sufficiently large electric fields were applied, domains of opposite polarity (with polarisation oriented into the surface plane) would nucleate and grow, traversing the interelectrode gap, to form conical (or needle-like) reverse domains [18-21], bounded by charged head-to-head domain walls. Previous work on the same kind of heterostructures [18] demonstrated that the subsurface inclination angles of the relatively mobile top section of walls could be monitored by mapping the locus of the domain wall on the LNO film surface. We used this technique to confirm the manner in which applied bias levels affected both the subsurface domain wall inclination behaviour and associated wall conductivity in our specific samples: after the initial application of negative bias (-1V) which was expected to align subsurface walls parallel to the polar axis, subsequent positive bias (+10V) caused domain wall traces (highlighted in red in figure 1c) to move by around ~10nm, in a sense that increased the fractional area occupied by the "downwardly" oriented conical domains. Assuming that the subsurface wall section length is ~50-100nm [18], this corresponds to a maximum of 11° change in the local subsurface wall inclination angle (figure 1d). The correlation between the inferred domain wall tilt variation and domain wall conductance was confirmed by cAFM current mapping at the domain wall using -1V and +6V tip bias (figure 1e). Diffuse current signals over the region in the vicinity of the domain wall (~100nm wide) are suspected to be due to both the finite size of the AFM tip (maintaining electrical contact with the domain wall over a range of tip positions) and a tendency for some current to traverse short lengths on the LNO surface before reaching high conductivity source-drain domain wall conduits.

**Diode behaviour and current rectification.** As expected from the bias-induced domain wall tilt variations and associated changes in conduction behaviour outlined above, clear diode-like current-voltage responses could readily be obtained in simple parallel-plate capacitor structures (using liquid GaInSn top electrodes and the thin film Cr/Au bottom electrode already present in the as-received thin film heterostructure) in which mixed domain states had been produced through partial poling (figure 2a). Such domain wall diodes were found to effectively enable ac current rectification, but only at low frequencies: for example, half-wave rectification output is clear at 1Hz (figure 2b). However, when the frequency is increased to 100 Hz, the response begins to show clear evidence of the superposition of domain wall transport and a capacitive current associated with the dielectric domains (figure 2c), acting electrically in parallel. When the frequency is increased to 1 MHz, rectification is completely lost (figure 2d); perhaps near-surface domain wall tilt variations are no longer able to respond to the relatively rapidly varying ac input voltages, and hence are no longer able to contribute to the overall ac current and so capacitive currents completely dominate. This would be consistent with the work of Schröder *et al.* [25], who made a similar observation on super-bandgap illuminated 300$\mu$m thick z-cut LNO parallel plate capacitors, where they found that the sample consisting of domain walls exhibited purely capacitive behaviour at frequencies greater than ~200 Hz.

**Diode-resistor logic gates.** Diode-resistor gates are a standard construction for the realisation of Boolean logic operations [26]. Figure 3a illustrates a typical diode-resistor potential divider arrangement that can be used to create "AND" gate binary output ($V_{out}$) from two binary inputs ($V_1$ and $V_2$). In our case, two partially switched LNO thin film capacitor structures (domain wall diodes), with thin film Ag top electrodes, were used along with a substantial 1G$\Omega$ resistor, placed between the diodes' positive terminals (upper electrodes) and the +5V supply; input signals were applied to the negative terminals of the domain wall diodes.

      When both inputs are set to logic "1" (+5V, figure 3c), there is no voltage drop in the circuit and an output voltage of ~+5V is measured (figure 3d). However, when one of the inputs is set to logic "0" (0V), a potential drop across the circuit is established. Since the "on" resistance of the diodes (diode resistance seen in forward direction after the barrier is surpassed) is on the order of M$\Omega$s (significantly lower than 1G$\Omega$) (figure S4), a relatively small fraction of the voltage is dropped across them and, as a consequence, $V_{out}$ registers between 1 and 2V (slightly above the barrier voltage) (figure 3d). Slight differences in the voltages recorded are due to the slight differences in the current voltage characteristics of the two domain wall diodes used. When both inputs to both diodes are set to logic "0" (0V), both act under forward bias and, since the resistance of the two diodes acting in parallel will be reduced further, a larger potential drop across the resistor develops and $V_{out}$ is reduced to ~1V. Provided the "0" logic output state is taken to be below ~2V, the logic output is "0" for all combinations of logic inputs, save for the case in which both inputs are "1" (see the truth table in figure 3f) and the "AND" gate is hence successfully demonstrated.

The diode-resistor circuit used to realise an "OR" logic gate is given in figure 3b. Here, the 1GΩ resistor is placed between the negative terminals of the two domain wall diodes and the ground, and input signals are applied to the positive diode terminals. The output ($V_{out}$) is the voltage dropped across the resistor. When both inputs ($V_1$ and $V_2$) are set to 0V (logic "0" states), there is no bias across the circuit as the potential is uniformly 0V at all points. Hence the measured $V_{out}$ is 0V (figure 3c and e). However, if either or both of the logic inputs are "1" (+5V), then the potential drop developed again acts in a forward bias sense across the diodes and their parallel resistance is relatively low. Most of the potential drop therefore occurs across the 1GΩ resistor and, since this is the output voltage, ~+3-4V is measured. The output logic state is hence "1" (provided it is defined as being above 3V) for all input logic combinations, save for the case when both inputs are "0" (see truth table figure 3g). A domain wall diode-enabled inclusive "OR" logic gate hence results.

The current-voltage characteristics of the two domain wall diodes associated with the logic gate data shown in figure 3 were explicitly measured (figure S4 in supplementary information) and the information used to develop predicted responses in two-level cascaded Boolean logic circuits, using the LTSpice analogue electronic circuit simulation software. We simulated two different scenarios (figure S5): "AND" gates in the first level followed by an "OR" gate in the second level (figure 4a), and "OR" gates in the first level followed by an "AND" gate in the second level (figure 4b). In both cases, generated output logic levels fell into the correct logic band (ideal logic levels are shown in figure 4d, e). Note, however, that in both cases, the voltage gap between defined logic states reduced significantly to <~1V. There are two reasons for this: firstly, the logic gate in the second stage experiences modified logic voltages generated by the first stage (figure 3d, e) at its input; secondly, the voltage generated at the output of the first stage is divided between the input resistance of the second stage and the output resistance of the first stage. This depresses the modified logic levels experienced by the second stage, so the gap between the logic levels at the output of the second stage is reduced further. Nevertheless, if states are defined appropriately, then the cascade logic using domain wall-enabled diodes is successfully demonstrated.

**Discussion and Outlook**

Diode-like behaviours have previously been observed in the characterisation of charge transport along domain walls, not only in LNO [18, 22, 27,28], but also in rare earth manganites such as $ErMnO_3$ [29]. However, these previously seen current-voltage asymmetries have almost entirely been associated with the contact resistance at the electrode-wall interface [29] and have hence been reflective of barrier physics, rather than being due to the inherent behavioural properties of domain walls themselves. Observations made by Zhang *et al.* [22] are perhaps the exception, as asymmetry in device conduction in their x-cut LNO coplanar capacitors appeared to be the result of making and breaking domain wall-electrode contacts under different senses of applied bias: one sense caused electrode-ferroelectric "interfacial" domains to contract to allow percolative domain wall pathways to be established between source and drain electrodes, while the other caused the "interfacial" domains to grow and hence break domain wall enabled source-drain electrical contact. It is interesting to note that, although this diode mechanism of making and breaking contact between domain walls and electrodes differs from that seen in our study, half-wave rectification was also only demonstrated at relatively low ac frequencies. This perhaps suggests that charged domain wall motion in LNO is generally rather sluggish.

Even if operationally slow, and similar in response to those investigated by Zhang *et al.* [22], the domain wall diodes developed in our work are conceptually unique. Rather than acting as mobile interconnects among arrays of fixed devices or contact points, our diodes operate because the walls themselves change their conductivity through localised changes in inclination angle; the walls therefore are the active devices that switch conductivity on and off internally. Such a notion is important for domain wall electronics development, as it shows that domain walls can simultaneously constitute both interconnects and devices; because walls can be created, moved and destroyed, this suggests the genuine possibility of fully ephemeral dynamic nanocircuitry [30]. The explicit demonstration that LNO domain walls, when acting electrically in parallel, can be used to realise Boolean logic gates takes this notion even further, showing that potentially transient domain wall circuits are capable of acting as information processing entities.

**Methods**

**Sample preparation:** A wafer with a 500 nm thick single crystal of undoped congruent z-cut lithium niobate (LNO), bonded to a 150 nm thick chromium-gold-chromium electrode stack, above a 2 μm silica layer, attached to a 500 μm thick z-cut lithium niobate, was commercially obtained from NanoLN. A piece of wafer was cut using a diamond scribe. In most functional measurements, 150 nm thick and ~100 x 100 μm$^2$ in area Ag electrodes were thermally evaporated on the top ferroelectric surface using a Moorfield MiniLab 060 type T60M system. The sample was then fixed on a printed circuit board, and its bottom chromium-gold-chromium electrode was contacted via conductive silver paint. 25 μm-diameter aluminium wire was wire bonded to Ag electrodes using a Kulicke & Soffa Model 4526 wirebonder.

**Scanning probe microscopy imaging:** An MFP-3D infinity AFM system from Asylum Research equipped with a high voltage conductive atomic force microscopy (HV-cAFM) holder was used for poling and piezoelectric force microscopy (PFM)/conductive atomic force microscopy (cAFM) experiments. A grounded platinum/iridium coated silicon AFM tip (Nanosensors model PPP-EFM) was used for the PFM/cAFM investigations. Scan rate was set ~3 Hz. For PFM scans, a drive amplitude of 10V was applied at ~340 kHz (tip resonance frequency). The domain structure probed in figure 1b, was created with a (50V, 1s) voltage pulse. After poling, a cAFM scan at -1V was performed to reset the inclination of subsurface domain walls. To obtain the associated domain wall positions, a PFM scan was carried out. Next, a cAFM scan at 10V was performed to enhance the inclination of near surface domain walls, and the associated conductance state was mapped with a 6V cAFM scan. Finally, a PFM scan was performed to extract associated domain wall positions. PFM images associated with reset and enhancement cAFM scans were superimposed (figure 1c). A spatial drift observed in the PFM images (due to the AFM drift) was corrected by aligning the centre points of them. All voltages denote the potential difference between top and bottom electrodes.

**Electrical measurements on macroscopic capacitor structures:** The wire-bonded samples were placed in an aluminium enclosure to protect them from electromagnetic noise. Electrical connections to the enclosure were achieved using triaxial cables.

**DC characterisation:** A Keysight B2910BL Source-Measure Unit was used for dc current-voltage measurements. Measurements (shown in figure 2a and figure S4) were performed under 100ms long voltage pulses. Delay, hold, and measurement times were set as 0ms, 80ms, and 20ms, respectively. 10 $\mu A$ of current compliance was set for all measurements.

**Current rectification:** An Agilent 33220A Arbitrary Waveform Generator was employed to generate sine waves in the 1 Hz – 1 MHz frequency regime (5V in amplitude), which were applied to sample's top electrode. A Rohde & Schwarz RTC 1002 digital oscilloscope was used to monitor the rectified ac voltages at the bottom electrode of the sample (figure S1).

**Logic gate experiment:** A Keysight E3631 single output dc power supply and a Keysight E3641 triple output dc power supply were used to generate the digital input signals shown in figure 3c. The current flowing through the resistor ($\sim 1\ G\Omega$) was measured with a Keysight B2910BL Source-Measure Unit which is connected in series ($V_{SMU}$ was set to 5 and 0V for the "AND" and "OR" gates, respectively) (figure S6a, b). $V_{out}$ was calculated as $5 - I_{Res}R$ and $I_{Res}R$ for the "AND" (figure S6a) and "OR" (figure S6b) gates, and plotted in figure 3d, e.

**Electronic circuit simulation:** LTSpice analogue electronic circuit simulation software was employed to simulate two-level logic implementations given in figure 4a, b. Firstly, two different diodes having the I-V characteristics shown in figure S4 ($D_1$ and $D_2$) were defined in LTSpice. Then, "AND" and "OR" logic gates were constructed using $D_1$, $D_2$ and resistors ($1\ G\Omega$) (figure S5). Next, logic gates were cascaded as shown in figure 4a, b. Periodic square waves $V_1$, $V_2$, $V_3$ and $V_4$ shown in figure 4c were generated and applied to the inputs of the cascaded gates. A transient analysis (160s long) was performed, and the generated output signals were plotted (figure 4d, e).

**Data availability**

The data that support the findings of this study are available from the corresponding authors upon reasonable request.

**Acknowledgements**

The authors are grateful for the financial and training support received from the EC H2020 "MANIC" Innovative Training Network (grant agreement no: 861153), the Engineering and Physical Sciences Research Council (EPSRC; through grant EP/P02453X/1 and through studentship funding) and the UKRI Future Leaders Fellowship programme (MR/T043172/1).

**Author contributions**

Most of the experiments and data analyses were done by AS. CMcC and JM helped in developing techniques to control the partial switching of the LNO samples. AK, RGPMcQ and JMG guided and supervised the research. JMG and AS were primarily responsible for conceptualisation of the work and for the initial drafting of the manuscript. All authors contributed to subsequent manuscript development.

**Competing interests**

The authors declare no competing interests.

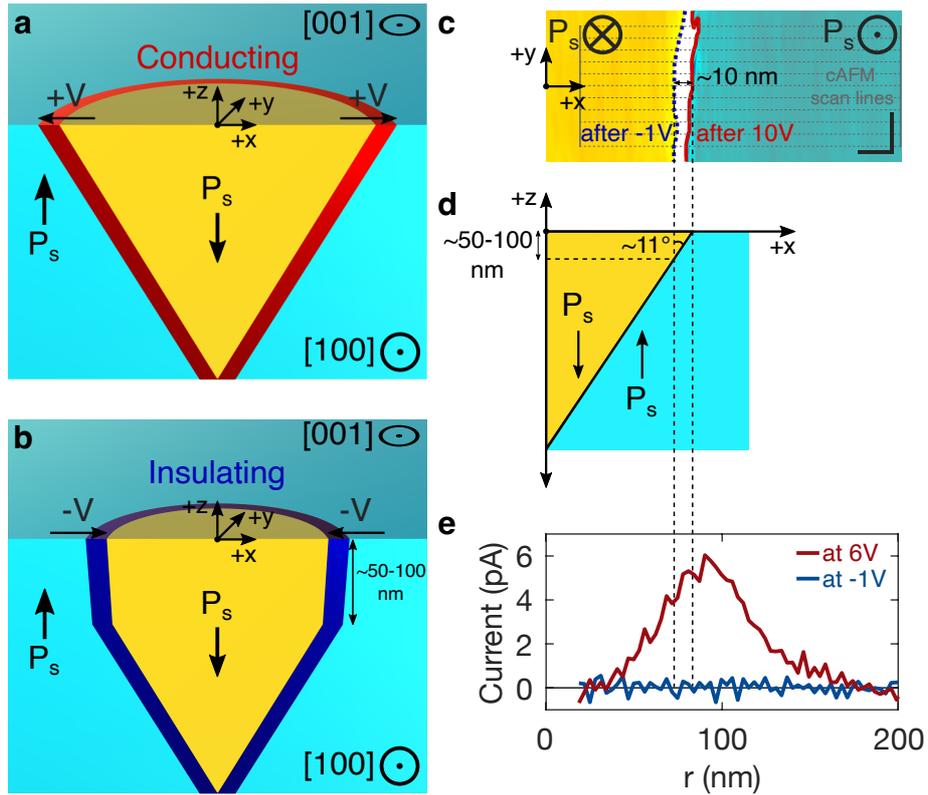

**Figure 1 | Electric field control of near surface domain wall inclination angle.** Domain walls near the top surface of the LNO thin film can be manipulated with modest applied bias: positive voltages tilt them away from the polar axis, leading to a "conducting" state (**a**), whereas negative voltages cause them to align parallel to the polar axis, resulting in an "insulating" state (**b**). This has been confirmed by piezoelectric force microscopy (PFM) (**c**) and conductive atomic force microscopy (cAFM) (**e**) investigations: domain walls (initially set into a low near surface inclination state by a -1V cAFM scan (highlighted in blue)) expand radially after a +10V cAFM scan (highlighted in red). A 10nm shift in domain wall position can result in up to 11° change in near surface domain wall inclination (depth of the subsurface domain wall region is estimated between 50-100nm [18]) (**d**). Enhanced current response is observed in cAFM data collected at +6V compared to -1V (**e**). Data shown is the average of the currents observed in the 11 cAFM scan lines highlighted in (**c**). Horizontal and vertical scale bars in (**c**) show 20nm and 5nm, respectively.

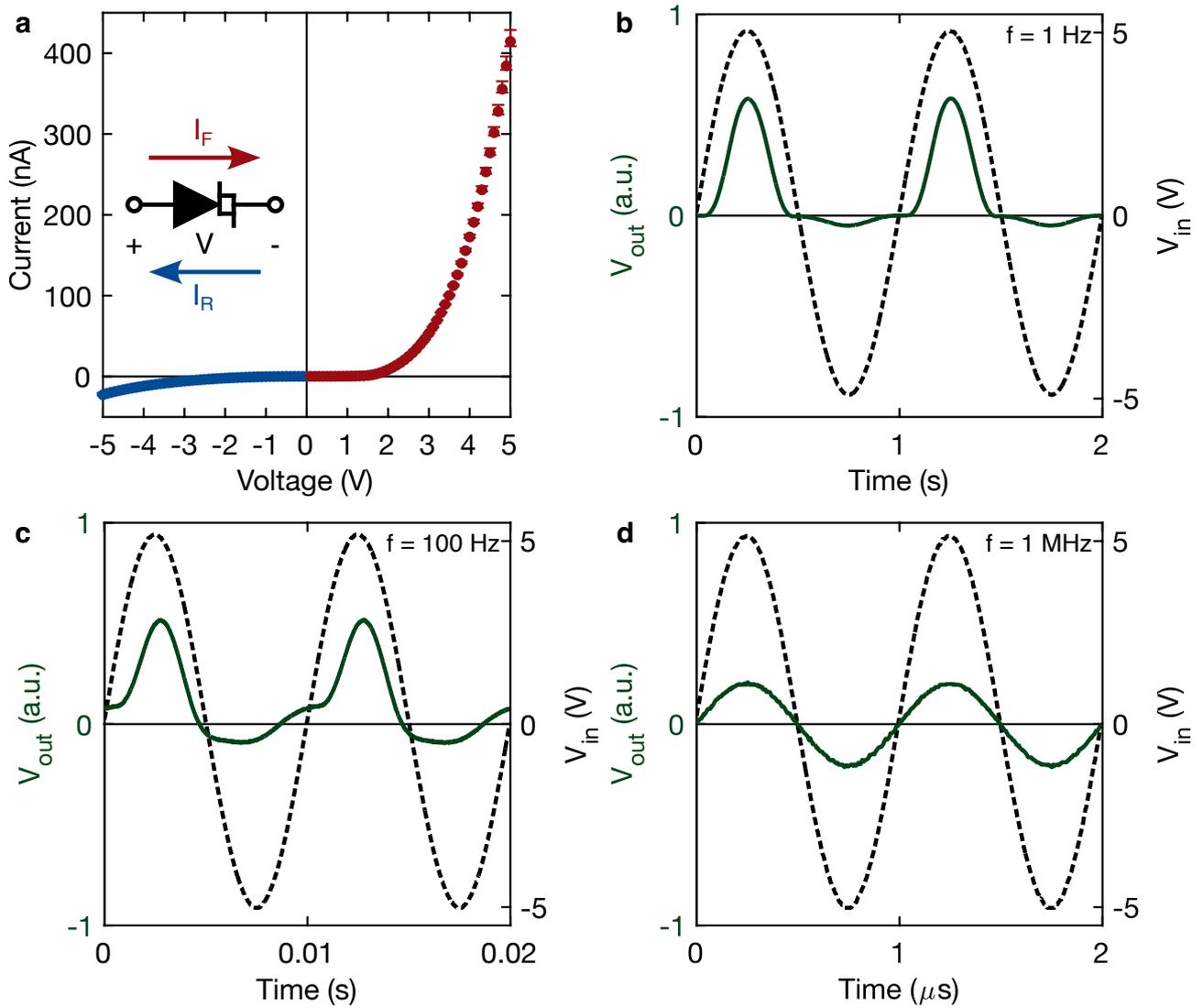

**Figure 2 | Diode-like behaviour in domain wall current and ac rectification.** Distinct asymmetry in the domain wall current-voltage characteristic is typical of diode-like behaviour (**a**). LNO thin film capacitor structures in which conducting domain walls are present can therefore be electrically modelled as a parallel combination of a diode and a capacitor (which is used to model the bulk capacitance) (figure S2). When a sinusoidal input voltage is applied at 1 Hz, the domain wall diode successfully rectifies the voltage across the load resistor ($V_{out}$) (**b**) according to the I-V characteristics shown in (**a**). In this case, the capacitor has little influence because of its extremely high impedance at this low frequency (**b**). When the input frequency is increased to 100 Hz, the capacitive impedance reduces, and the output voltage waveform becomes distorted (**c**). At 1 MHz of input frequency, output current becomes fully capacitive, which is inferred by the fact that output voltage becomes almost fully symmetric about x-axis. This implies that asymmetric domain wall current response has little influence at 1 MHz (**d**).

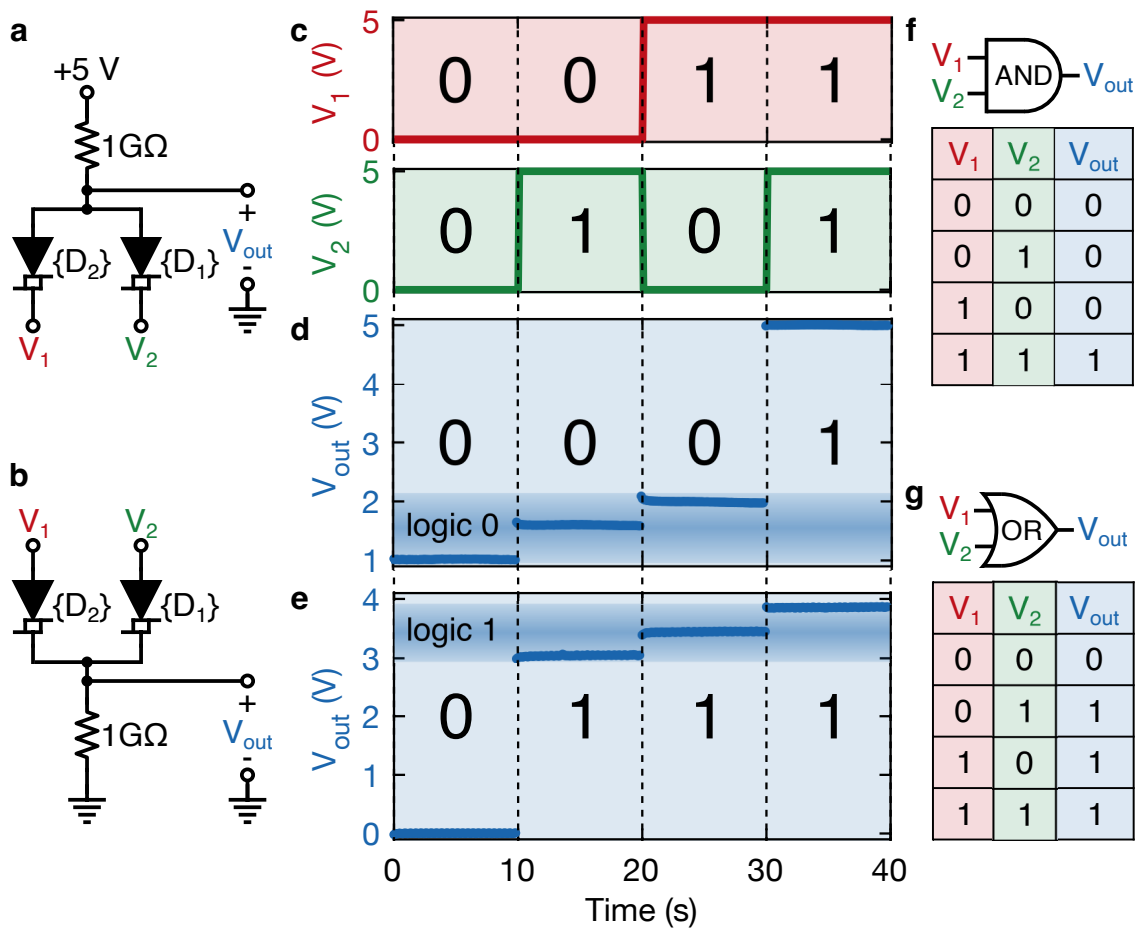

**Figure 3 | Logic gate implementations based on two domain wall diodes and a resistor.** Circuit configurations shown in (**a**) and (**b**), consisting of two domain wall diodes and a resistor, realise "AND" and inclusive "OR" gates, respectively. For the "AND" gate shown in (**a**), when both inputs are set to logic "1" (+5 V) (**c**), output becomes logic "1", otherwise it is set to logic "0", which is defined by the blue-coloured voltage band in (**d**). For the "OR" gate shown in (**b**), when either $V_1$ or $V_2$ or both are set to logic "1" (+5 V) (**c**), the output generated is logic "1", defined by the blue-coloured band (**e**); otherwise it is set to logic "0" (0 V). Input-output relationship of "AND" (**f**) and "OR" (**g**) gates are demonstrated in truth tables.

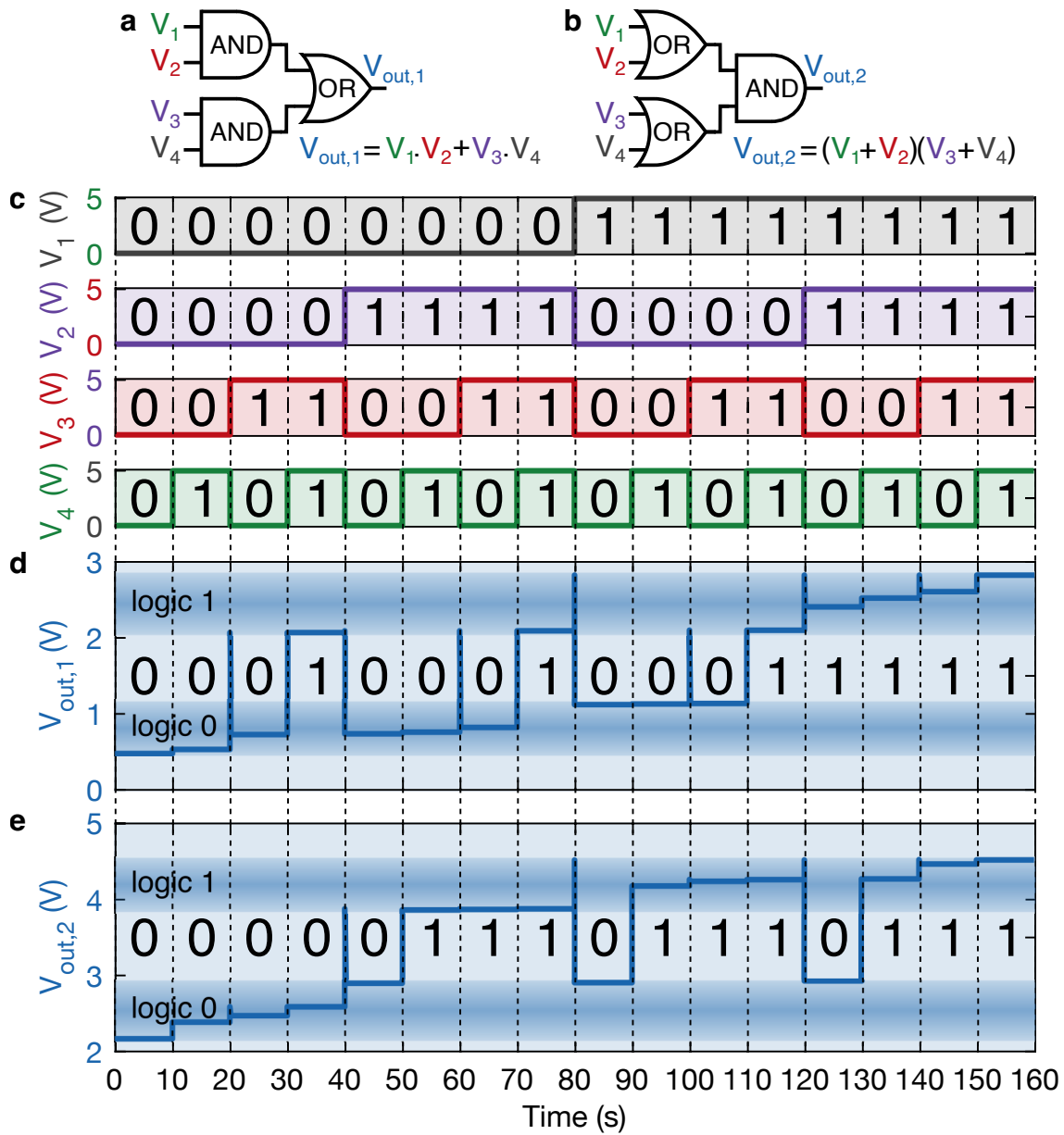

**Figure 4 | Simulation of two-level logic based on domain wall logic gates.** Two different examples of two-level Boolean logic are simulated using the measured current-voltage characteristics of the two domain wall diodes used to obtain data shown in figure 3: "AND" gates in the first layer followed by an "OR" gate (**a**) and "OR" gates in the first layer followed by an "AND" gate (**b**). All possible logic input combinations are applied through time (**c**), and the output voltage is simultaneously recorded (**d**, **e**). For both cases, the correct output logic levels are successfully maintained in output voltages with ~1V of separation between "1" and "0".